# $Yb_4LiGe_4$ – A Yb Mixed Valent Zintl Phase with Strong Electronic Correlations

Sebastian C. Peter[1*], Steven M. Disseler[2], J. Niclas Svensson[2], Pietro Carretta[3], Michael J. Graf[2]

[1]New Chemistry Unit, Jawaharlal Nehru Centre for Advanced Scientific Research, Jakkur, Bangalore 560064, India
[2]Department of Physics, Boston College, Chestnut Hill, MA 02467 USA
[3]Department of Physics "A.Volta" and CNISM, University of Pavia, 27100 Pavia, Italy

**Abstract**

Single-phase samples of $Yb_4LiGe_4$ and $Yb_5Ge_4$ were synthesized using high frequency (HF) heat treatment. $Yb_4LiGe_4$ crystallizes in orthorhombic space group *Pnma* with the $Gd_5Si_4$ type of crystal structure and lattice parameters $a = 7.0571(1)$ Å, $b = 14.6239(1)$ Å, and $c = 7.6155(1)$ Å. One Yb position in $Yb_5Ge_4$ is substituted by the lithium atom and causes a distortion of the germanium tetragons in $Yb_4LiGe_4$. Investigation of the electronic state of Yb via magnetic susceptibility and X-ray absorption near-edge spectroscopy (XANES) revealed a presence of two electronic states of ytterbium, $4f^{13}$ and $4f^{14}$ (mixed valence), in $Yb_5Ge_4$ and $Yb_4LiGe_4$. Studies of the temperature dependence of the electrical resistivity, magnetization, $^7Li$ spin-lattice relaxation rate and the specific heat indicate that strong electronic correlations are present in $Yb_4LiGe_4$, and below approximately 50 K there is a competition between ferromagnetic and antiferromagnetic correlations. Magnetic ordering in $Yb_4LiGe_4$, if present, occurs below the reported antiferromagnetic transition temperature of 1.7 K for $Yb_5Ge_4$.

PACS Numbers: 71.20.LP, 71.27.+a, 71.2d.+d

*Corresponding author. Phone: 080-22082298, Fax: 080-22082627
sebastiancp@jncasr.ac.in (S. C. Peter)

1. **Introduction**

Compounds with a general formula $RE_5Tt_4$ ($RE$ = Rare-earth; Tt = Si, Ge) have been known for decades [1]. However, recently these materials captured the intense interest of the scientific community after the discovery of the giant magnetocaloric, magnetostriction and magnetoresistance effects in the $Gd_5Si_4$-$Gd_5Ge_4$ system [2-9]. Pecharsky and co-workers demonstrated that the above mentioned physical properties can be tuned by the substitution of Ge with the isoelectronic Si in the $Gd_5(Si_xGe_{1-x})_4$ system [2-5, 7-9]. The family of $Gd_5Si_xGe_{4-x}$ alloys demonstrates a variety of unique physical phenomena related to magneto-structural transitions associated with reversible breaking and reforming of the interslab $T_2$ dimers that can be controlled by numerous external parameters such as chemical composition, magnetic field, temperature, and pressure. Similarly, the partial substitution of Ge by electron poorer Ga changes the valence electron count and affects the bond-making/breaking processes concomitant with the magnetic ordering [10]. These and other possibilities for exchanging Tt elements have been explored in the past. Interestingly, a different approach to tuning the material properties has been the substitution of the lanthanide metal with non magnetic element within the $Gd_5Si_4$ type structure, for example in $La_{5-x}Ca_xGe_4$ [11], $Ce_{5-x}Ca_xGe_4$ [11] and $Yb_{5-x}Mg_xGe_4$ [12] compounds.

Since Li is diagonally related to Mg and $Yb_5Ge_4$ exhibits mixed valent behavior, there exists the possibility for interchanging Li for either trivalent or divalent Yb. We have synthesized $Yb_{5-x}Li_xGe_4$ with the assumption that the compound may show a probable structural transformation and possibly interesting new physical properties. Several systems combining rare-earth elements with germanium and lithium have been studied [13-31]. Among these, the Yb-containing compounds have particular scientific interest because they can exhibit two energetically similar electronic configurations: the magnetic $Yb^{3+}$ ($4f^{13}$) and the nonmagnetic $Yb^{2+}$ ($4f^{14}$) one. Due to this feature Yb is usually considered as the "$f$-hole" analogue of Ce. In this case, the roles of the $4f$ electron and $4f$ hole can be interchanged, and many phenomena, such as intermediate valence, Kondo effect or heavy-fermion behavior, are observed in Ce and Yb analogues [32-35].

The crystal structure of $RE_4LiGe_4$ ($RE$ = rare-earth metals) ternary compounds have been reported by *Pavlyuk et al* from the X-ray powder diffraction (XRPD) data [21]. A full crystal structure investigation has been performed from powder diffraction data for $Tm_4LiGe_4$ [21] and for $Yb_4LiGe_4$ ($Yb_2Li_{0.5}Ge_2$) using single crystal diffraction data [28]. Properties of $Yb_4LiGe_4$ have not yet reported.

In the paper, we report the synthesis of single-phase $Yb_4LiGe_4$ and $Yb_5Ge_4$ samples using a high frequency (HF) heat treatment method. $Yb_5Ge_4$ is a mixed valent system adopts $Gd_5Si_4$ type structure in orthorhombic space group *Pnma*. $Yb_4LiGe_4$ is a ternary variant of $Yb_5Ge_4$ substituting Li at one of the Yb position of $Yb_5Ge_4$ which further enhances the average valence of the Yb. Using high pressure we produced dense samples with well-defined dimensions and shape for measurements of physical properties like electrical and thermal transport and specific heat capacity. Because the band structure is highly sensitive to changes in the chemical bonding, we have carried out a broad characterization of electronic properties in the temperature range 1.8 K to 350 K employing X-ray absorption spectroscopy, heat capacity, DC and AC magnetic susceptibility, electrical resistivity, and NMR spectra and relaxation measurements. As shall be shown, these results provide clear evidence that $Yb_4LiGe_4$ is a strongly correlated system with competing magnetic interactions.

2. **Experimental**

The starting materials for the preparation of $Yb_5Ge_4$ and $Yb_4LiGe_4$ were ingots of ytterbium (99.99 wt%, Lamprecht), lithium rods (99.4 wt%, Alfa Aesar), and germanium pieces (99.9999 wt%, Chempur). The elements in stoichiometric ratios were heated in closed tantalum tubes to roughly 1070 K using a high frequency furnace. Careful handling of the ytterbium and lithium was necessary in order to minimize impurities, and was carried out inside an argon glove-box system. The synthesized samples were loaded into a planetary ball mill (Pulverisette 7 Classic, Fritsch GmbH, Germany) for milling of about half an hour at a rotation speed of 600-800 rpm. In order to avoid excess heating of the vessels during milling, the process included equal pauses followed by active periods. The XRPD patterns of the samples obtained before and after milling were identical. Powder blends obtained from the milling process were then loaded again into Ta tubes and repeat the above process.

Phase analysis of both compounds was done by powder X-ray diffraction using a Guinier Huber G670 imaging plate camera applying Cu K$_{\alpha 1}$ radiation ($\lambda$ = 1.54056 Å). The patterns were used to obtain the unit cell parameters by a least-squares refinement of the positions of the lines, calibrated by lanthanum hexaboride as an internal standard ($a$ = 4.1569 Å).

Semi-quantitative microprobe analyses of the samples were performed with a Hitachi S-3400 scanning electron microscope (SEM) equipped with a PGT energy dispersive X-ray analyzer. Data were acquired with an accelerating voltage of 20 kV and a 60 s accumulation time. The EDS analysis taken on visibly clean surfaces of the samples gave the atomic composition close to 5:4 and 4:4 for $Yb_5Ge_4$ and $Yb_4LiGe_4$, respectively.

The $L_{III}$ X-ray absorption spectra (XAS) of $Yb_5Ge_4$ and $Yb_4LiGe_4$ were recorded in transmission arrangement at the EXAFS II beamline E4 of HASYLAB at DESY. Wavelength selection was realized by means of a Si (111) double crystal monochromator which yields an experimental resolution of approximately 2 eV (FWHM) for the experimental setup at the Yb $L_{III}$ threshold of 8942 eV, and experiments were carried out in the energy range of 8829 eV to 9180 eV. Experimental data were measured using $Yb_2O_3$ as an external reference for the $4f^{13}$ configuration. Deconvolution of the XAS spectra was made by the program XASWin [36].

Measurements of the DC magnetization in applied magnetic fields between 10 mT and 7 T and in the temperature range 1.8 K - 400 K were carried out with a SQUID magnetometer (MPMS XL-7, Quantum Design). In addition, AC susceptibility measurements were made using the Oxford MagLab platform on a second batch of $Yb_4LiGe_4$ samples, and confirmed the DC magnetization results presented here. Samples used for magnetic characterization were cut from disks produced by hot pressing of powder originally produced in a bulk form. Temperature-dependent magnetic susceptibility measurements of $Yb_4LiGe_4$ at 0.1 T show no difference from the powdered material, indicating a negligible inclusion of magnetic impurity phases.

The electrical resistivity measurements were performed on the same sample batch employing both DC and AC four-point methods in the temperature range 1.8 K - 300 K. The heat capacity was measured for 3 K < T < 60 K and in magnetic fields to 3 T using in the Oxford MagLab semi-adiabatic relaxation calorimeter platform. The heat capacity of the empty platform, which is less than 5% of the total heat capacity over the entire temperature range studied, was independently measured and subtracted off to obtain the sample specific heat.

$^7Li$ NMR measurements were carried out on a TecMag Apollo spectrometer by using standard radiofrequency pulse sequences. The NMR spectra were obtained from the Fourier transform of half of the echo after a $\pi/2$ - $\tau$ - $\pi$ pulse sequence. The characteristic decay time of the echo $T_2$ was measured at 300 K and 90 K, and found equal to 475 $\mu s$ and 207 $\mu s$, respectively. The spin-lattice relaxation rate $1/T_1$ was determined by following the recovery of nuclear magnetization after a saturating sequence of $\pi/2$ pulses. Data were taken in an applied magnetic field of 10 kG, with several additional points at 5 kG to demonstrate the relative insensitivity of the results to the field strength. The recovery law for the nuclear magnetization was found to be of the form

$$M = M_0 \left(1 - \exp\left[-(t/T_1)^\beta\right]\right),$$

that is, a stretched exponential, with a nearly temperature-independent exponent of $\beta = 0.9$. The deviation from a conventional exponential recovery law is likely to be due to the hyperfine coupling tensor anisotropy.

## 3. Results and discussion

### 3.1. Preparation

The $Yb_4LiGe_4$ sample after HF treatment is polycrystalline in nature, light grey with metallic luster and slightly air sensitive. The powder X-ray pattern of the first step synthesis reveals the presence of secondary phases ($Yb_3Ge_{4.3}$, Yb and Ge) in addition to $Yb_4LiGe_4$. The main additional phase was identified as orthorhombic compound $YbLiGe_2$ [30] and was confirmed by EDX. The not completely reacted ytterbium and germanium may be present as elements and/or in form of $Yb_3Ge_{4.3}$, and may not have been detected by XRPD because of the complexity of the diffraction pattern. The first step material was then processed again by HF treatment and the resulting product was checked with X-ray diffraction and metallography. Only a single phase $Yb_4LiGe_4$ was observed within the detection limit of the X-ray diffractometer. The optical microscopy and the EDX results confirm the quality of the sample and reveal negligible amounts of elemental Ge as a secondary phase.

Single phase $Yb_5Ge_4$ was also synthesized by two step process. Initially we used a similar synthesis procedure as reported by Ahn *et al* [37]. The X-ray powder diffraction pattern of $Yb_5Ge_4$ sample after the first HF treatment showed mainly reflections of the orthorhombic crystal structure of the $Gd_5Si_4$ type. Small amounts of $Yb_5Ge_3$ and Ge were observed in the XRPD and in metallographic study. However, the sample obtained after second HF treatment was completely pure $Yb_5Ge_4$ as determined by XRPD.

The Guinier X-ray powder diffraction patterns were used to refine the unit cell parameters using lanthanum hexaboride as internal standard. The powder diffraction patterns of $Yb_5Ge_4$ and $Yb_4LiGe_4$ were indexed on the basis of the orthorhombic primitive cell with the space group *Pnma*. The lattice parameters are listed in Table 1. Whereas the lattice parameters of $Yb_5Ge_4$ obtained in the present investigation are equal to the literature values within a few e.s.d. [37-40] the cell dimensions for our lithium-containing samples show some significant differences from the different literature data [37-40]. In all three cases, the lattice volume is smaller than that of $Yb_5Ge_4$ suggesting the successful implementation of Li into the initial crystal structure of the $Gd_5Si_4$ type. The marked variation of the lattice parameters in different studies indicates the differing levels of the Yb-by-Li substitution and, in fact, formation of the solid solution of Li in the binary ytterbium germanide - $Yb_{5-x}Li_xGe_4$. Assuming that the Li content in the sample investigated in Ref. 28 agrees with $x = 1$, and applying Vegard's rule on the lattice volume, we conclude that the Li content in our sample is slightly smaller than 1 ($x \approx 0.94$).

### 3.2. Crystal structure

The simplest description of the parent compound $Yb_5Ge_4$ is given by the formula $[Yb_5]^{12+}[Ge_2]^{6-}[Ge_2]^{6-}$, and detailed studies of $Yb_5Ge_4$ [37-40] revealed mixed-valence behavior of ytterbium, and the formula can be written as $[Yb^{2+}]_{5-x}[Yb^{3+}]_x[Ge_2]^{6-}[Ge_2]^{6-}$. One position (Yb1) shows notably larger distances to the germanium ligands (first four shortest distances in the coordination sphere are 3.06, 3.14, 3.16 and 3.17 Å) if compared to the Yb2 position (the first four shortest distances are 2.86, 2.92, 2.92 and 2.93 Å). The third position (Yb3) reveals intermediate coordination behavior (2.96, 2.98, 2.98 and 3.17 Å) being rather closer to Yb2. The first site may be considered as a suitable for $Yb^{2+}$ species, while the two others may support $Yb^{3+}$ species. Yb1, Yb2 and Yb3 in $Yb_5Ge_4$ occupy at 8$d$, 8$d$ and 4$c$ Wyckoff positions, respectively. Assuming that the formal charges for the germanium dumb-bells are correct, this leads to the formula $[Yb^{2+}]_3[Yb^{3+}]_2[Ge_2]^{6-}[Ge_2]^{6-}$ [28]. These features open the possibility of changing of the electronic state of ytterbium and approaching the electron balance state by the substitution of the Yb atom in particular by a element with the lower formal charge (valence). One example of such a chemical modification by substitution of ytterbium by magnesium - $Yb_{5-x}Mg_xGe_4$ - was recently reported [12]. $Yb_4MgGe_4$ has alternating layers of $Yb_2MgGe_2$ and $Yb_2Ge_2$ along the b-direction, with a charge arrangement approximated by $(Yb^{2+})_2Mg^{2+}(Ge_2^{6-})$ and $(Yb^{3+})_2(Ge_2^{6-})$ following the Zintl Klemm concept [12, 28]. A consequence of substituting Li for Mg in this structure is that to maintain adequate charge balance, there should be either an increase in the average valence of the Yb in the $Yb_2LiGe_2$ psuedo-plane, or a slight undercharging of the Ge dimers leading to bond instabilities; our magnetic susceptibility measurements, presented below, indicate that the former explanation is correct. Here, magnesium substitution was observed on two positions Yb2 and Yb3, suggesting that the size but not the electronic factor is playing a more important role in this substitution chemistry. Using their crystal chemical analogy, lithium can be also used for replacement instead of magnesium. The presence of Li in the crystal structure was proved by Xie *et al* in their single crystal X-ray diffraction studies [28]. Since we haven't succeed to grow single crystals, we refined the crystal structure of $Yb_4LiGe_4$ from the powder diffraction data using full profile method [41], shown in Figure 1. The atomic coordinates obtained from our refinement are in very good agreement with the reported structure [28] (Tables 2 and 3). In agreement with the single crystal data, $Yb_4LiGe_4$ shows lithium substitution on the Yb3 position which has lower coordination number, and, thus, is suitable for the smaller lithium atoms. This finding reveals that the size ratio of the substituting atoms is more relevant for the formation of the solid solution $Yb_{5-x}Li_xGe_4$, but not the formal charges derived from the interatomic distances as described above. It is also worth noting that because Li substitutes for Yb at the Yb3 position, this reduces the magnetic coupling between $Yb^{3+}$ layers and potentially reducing the dimensionality of the system.

In order to shed light on the influence of lithium on the crystal structure of $Yb_5Ge_4$ it is worth comparing the different 2D segments in $Yb_5Ge_4$ and $Yb_4LiGe_4$ for $0.05 < y < 0.45$ (segment 1, Figures 2a and 2b) and for $-0.22 < y < 0.22$ (segment 2, Figures 2c and 2d). Segment 1 resembles the atomic arrangement in the structural motif of the FeB type (in the binary compound it has a composition '$Yb_2Ge_2$'), and segment 2 shows a pattern of the $U_3Si_2$ type (composition 'Yb$_3$Ge$_2$'). The total composition may be obtained as Yb$_2$Ge$_2$ + Yb$_3$Ge$_2$ = Yb$_5$Ge$_4$. Different segments contain different [Ge$_2$]$^{n-}$ species: segment 1 bears [Ge$_2$]$^{6-}$ with a smaller Ge-Ge distance, segment 2 includes [Ge$_2$]$^{6-}$ with a larger Ge-Ge distance. The lattice parameters of Yb$_5$Ge$_4$ are much larger than that of Yb$_4$LiGe$_4$ (see Table 2). This, of course, can be related to the fact that Yb is larger than Li. On the other hand, this leads to a slightly shorter Ge-Ge distances in Yb$_4$LiGe$_4$ (2.54 Å in segment 1 and 2.60 Å in segment 2) compared to those in Yb$_5$Ge$_4$ (2.57 Å in segment 1 and 2.65 Å in segment 2). Another striking feature after substituting the Yb3 position with Li atom is the difference in angle between the Ge pairs in different segments. It remains very close in segment 2 with 32.11º and 32.75º for Yb$_5$Ge$_4$ and Yb$_4$LiGe$_4$, respectively. However, segment 1 shows a remarkable distortion after the substitution as it increased from 34.63º (Yb$_5$Ge$_4$) to 37.51º (Yb$_4$LiGe$_4$).

### 3.3. X-ray absorption spectroscopy (XAS)

The Yb-$L_{III}$ spectra for Yb$_4$LiGe$_4$ and Yb$_5$Ge$_4$ (Figure 3) were fitted by XASWin[36] and display two peaks: one at 8948 eV associated with Yb$^{3+}$ and the other at 8941 eV associated with Yb$^{2+}$. The spectra of both compounds confirm the finding of mixed valence behavior. The evaluated average valence of the Yb atom is 2.57 in Yb$_4$LiGe$_4$ and 2.42 in Yb$_5$Ge$_4$, consistent with the mixed-valent state as inferred for Yb$_5$Ge$_4$ [37, 40]. The increase of the average valence of Yb with the lithium substitution is consistent with the electron balance formulation explained above, and confirmed by magnetic measurements, as described in Section 3.5.

### 3.4. Heat Capacity

In Figure 4 we show our results for the specific heat of Yb$_4$LiGe$_4$. The low-temperature increase and plateau near 3 K (shown in inset) are similar to the behavior observed in the parent compound Yb$_5$Ge$_4$ [37], and are attributed to magnetic degrees of freedom. Application of a magnetic field of 3 T has very little effect on the observed behavior. The unusual temperature dependence will be discussed in Section 3.7 in the context of critical fluctuations as precursors to a magnetic phase transition.

### 3.5. Magnetic Susceptibility

At temperatures above 60 K the susceptibility of $Yb_4LiGe_4$ exhibits paramagnetic Curie-Weiss behavior $\chi = C/(T-\theta_P)$, where $\theta_P$ is Curie-Weiss temperature and $C$ is a constant, as shown in Figure 5. From curve fits of the inverse susceptibility versus temperature we extracted a negative $\theta_p = -3(1)$ K, compared to -4 K reported for $Yb_5Ge_4$ [37] and consistent with antiferromagnetic correlations. The slope yielded an effective magnetic moment of 3.71(1) $\mu_B$ per Yb, larger than the 2.72 $\mu_B$ per Yb reported for $Yb_5Ge_4$ [37]. Following the analysis from Ref. 37, the fraction of Yb sites occupied by each valence state can be estimated from

$$\mu_{eff} = [n_{2+} (\mu_{2+})^2 + (1- n_{2+}) (\mu_{3+})^2]^{1/2},$$

where $\mu_{2+}$ and $\mu_{3+}$ are the theoretical free magnetic moments of each valence state (0 $\mu_B$ and 4.54 $\mu_B$, respectively), and $n_{2+}$ is the fraction of Yb sites occupied by the non-magnetic $Yb^{2+}$. The result is an average value of $n_{2+} = 0.33(5)$.

First, we note that the estimated fraction of magnetic $Yb^{3+}$ sites for $Yb_4LiGe_4$ $n_{3+} = 0.66$ is slightly larger than the 0.57 value extracted from the XAS measurements, and significantly larger than the value 0.36 reported for $Yb_5Ge_4$ [37]. This result is contrary to the expectation that Li would preferentially substitute for the smaller $Yb^{3+}$ at the crystallographic 4$c$ site, an inference consistent the quantum chemical calculations of Zhang and Miller [42] for $Yb_4MgGe_4$. Rather, these results are again consistent with the proposal that an excess of electrons is required to stabilize the material, resulting in an increase in the average valency for the Yb ions.

Finally, we note that the Yb ions in $Yb_5Ge_4$ form a Shastry-Sutherland type lattice [43], as can be seen by comparing our Fig. 2b to Fig. 5a of Ref. 43, and so geometric frustration effects could also play a role in the observed effective magnetic moment. It is interesting to speculate whether the lattice distortion created by Li substitution could in fact modify the degree of frustration thereby influencing the observed moment.

### 3.6. Electrical Resistivity

The room temperature resistivity is fairly large, roughly 1500 $\mu\Omega$-cm, one to three orders of magnitude larger than the resistivity of 'typical' Yb-based intermetallic compounds. However, the sample remains a conductor down to 2 K, consistent with band structure calculations [28]. As shown in Figure 6, the resistivity is seen to increase sharply with decreasing temperature, but exhibits a broad maximum near 100 K, then increases very weakly upon further cooling. The behavior is qualitatively similar to that observed in strongly correlated Yb-systems with competing magnetic interactions [44, 45]. The inset to the figure shows a second small maximum near 10 K. Since no other properties show a structure at this temperature, which is well above the reported magnetic ordering temperature for $Yb_5Ge_4$, we suggest, that this feature reflects a second energy scale to the electronic correlations in the system. However, we cannot rule out the possibility that the Yb oxidation states are varying with temperature, which could also contribute to the observed resistivity versus temperature curve.

### 3.7. $^7$Li Solid State NMR

The temperature dependence of the spin-lattice relaxation rate $1/T_1$ (Figure 7) shows an unusual non-monotonic behavior, with $1/T_1$ increasing as temperature is lowered from 300 K, reaching a broad maximum located roughly near 60 K, and decreasing before attaining a nearly constant level near 4 K. This behavior is not typical of simple metals or of localized moment systems, but rather resembles behavior seen, for example, in the strongly correlated system the $CeCu_{6-x}Au_x$ system, where Kondo screening competes with the tendency towards magnetic ordering [46].

Fourier transforms of the $T_1$ pulse echo reveal a narrow line width of 10 kHz at 300 K, and which increases to about 250 kHz at 4 K; the low-temperature spectrum was confirmed using a point-by-point method. The line width varies nearly linearly with dc susceptibility $\chi$ (Figure 8) over the temperature range 300 K to 20 K. For a polycrystalline sample such as ours, the anisotropic hyperfine coupling will produce a distribution of resonance frequencies, and from the slope we estimate this anisotropic hyperfine coupling to be approximately 650 G, typical of dipolar coupling. It is interesting to notice that below about 20 K a deviation from the linear trend shown in Figure 8 is observed. This suggests a modification in the hyperfine coupling which might be associated with a change in the band structure. Assuming localized spins, the resulting exchange coupling is of order 5 K - 10 K, a value too large to be associated with purely dipolar coupling, and may be indicative rather of RKKY coupling of the $Yb^{3+}$ ions. On the other hand, if we assume instead delocalized spins, 10 K represents the energy scale of correlations in the system. This latter interpretation is consistent with the observed maximum in the temperature-dependent resistivity and spin-lattice relaxation data. The results reported above show that $Yb_4LiGe_4$ share many properties with the Yb-based intermetallic compounds which are proximate to a magnetic ordering transition and can be tuned via substitution or pressure to exhibit non-Fermi liquid behavior [44, 45]. Following Moriya's theory for quantum fluctuations in the vicinity of a low-temperature magnetic transition [47], in Figure 9a we plot $1/T_1T$ versus dc susceptibility $\chi^{3/2}$ in the low temperature range $T < 30$ K, and find a linear variation; according to Ref. 38 this behavior is consistent with the presence of two-dimensional *ferromagnetic* fluctuations. In Figure 9b we plot $C/T$ of $Yb_4LiGe_4$ versus $T^{-1/3}$ and find a linear behavior below 10 K, which is again consistent with Moriya's theory for a system exhibiting two-dimensional ferromagnetic fluctuations. This result seems highly unusual. While it is possible that the layered structure [12] of the material could produce two dimensional behavior (especially in light of the Li substitution at the Yb3 lattice position), the parent compound is reported to undergo *antiferromagnetic* ordering, consistent with the negative Curie temperature extracted from our high temperature susceptibility results. A crossover from antiferromagnetic to ferromagnetic correlations has been also reported for Ce compounds on a Shastry-Sutherland lattice [43]. This crossover could occur around 10 K - 20 K where the change in the hyperfine coupling and the upturn in the resistivity are observed.

The inferred competition between Kondo screening of the local moments and RKKY interactions driving $Yb_4LiGe_4$ towards magnetic order, as well as the possible effects of magnetic frustration, make this material a likely candidate to exhibit unusual behaviors typically associated with competing order parameters, e.g., heavy fermion or quantum critical behavior. We are currently extending our measurements below 1.8 K to look for signatures of a strongly-correlated ground state, as well as to determine if the unusual critical behavior discussed above is related to a magnetic phase transition.

## 4. Concluding remarks

Single phase of $Yb_5Ge_4$ and $Yb_4LiGe_4$ were obtained by high frequency induction heating. Substitution of Li at Yb position in the $Yb_2LiGe_2$ psuedo-plane of parent compound $Yb_5Ge_4$ enhances the average valence of the Yb. X-ray absorption spectra in combination with the measurements of electrical conductivity and magnetic susceptibility on $Yb_4LiGe_4$ show presence of both $Yb^{2+}$ and $Yb^{3+}$ Measurements of the electronic transport, nuclear spin-lattice relaxation, specific heat, and magnetization all point to the existence competing electronic interactions with multiple energy scales, and that the ground state may not be a simple antiferromagnet, as reported for the parent system $Yb_5Ge_4$. The lack of conventional magnetic order would be consistent with the geometric frustration expected for the structure of the Yb sub-lattice, and with the mixed valence of the Yb ions.

**Acknowledgements**

This work was supported in part by National Science Foundation – Materials World Network grant DMR-0710525. The authors express their gratitude to the facilities at the EXAFS II beamline E4 of HASYLAB at DESY and U. Burkhardt for measurements.

**Supporting Information Available**. Powder XRD pattern of $Yb_5G_4$ and $Yb_4LiGe_4$.

**Tables**

Table 1. Lattice parameters of $Yb_{5-x}Li_xGe_4$ (x = 0, 1).

| Composition | $a$, Å | $b$, Å | $c$, Å | $V$, Å$^3$ | Reference |
|---|---|---|---|---|---|
| $Yb_5Ge_4$ | 7.3430(9) | 14.959(2) | 7.829(1) | 859.9(4) | this work |
| $Yb_5Ge_4$ | 7.342(2) | 14.958(1) | 7.828(1) | 859.7 | 39 |
| $Yb_5Ge_4$ | 7.3406(5) | 14.9423(9) | 7.8253(5) | 858.3 | 37 |
| $Yb_4LiGe_4$ | 7.0828(3) | 14.6415(7) | 7.6279(4) | 791.0(3) | this work |
| $Yb_4LiGe_4$ | 7.0601(6) | 14.628(1) | 7.6160(7) | 786.5 | 28 |
| $Yb_4LiGe_4$ | 7.01 | 14.41 | 7.53 | 760.6 | 21 |

Table 2. Crystallographic data for Yb$_4$LiGe$_4$ obtained from this work compared with the reported singe crystal data.

|  | Powder (this work) | Single Crystal |
|---|---|---|
| Mode of refinement | Full profile | Full matrix least squares on F$^2$ |
| Space group, Z | *Pnma*, 4 | *Pnma*, 4 |
| Radiation, $\lambda$ (Å) | Cu *K$\alpha$*, 1.5406 | Mo *K$\alpha$*, 0.71073 |
| Unit cell parameters | | |
|     *a* (Å) | 7.0828(3) | 7.060(1) |
|     *b* (Å) | 14.6415(7) | 14.628(1) |
|     *c* (Å) | 7.6279(4) | 7.616(1) |
|     *V* (Å$^3$) | 791.0(1) | 786.5(1) |
| $\rho_{calcd}$ (g/cm$^3$) | 8.308 | 8.356 |
| Reflections/parameters | 1644/23 | 1599/44 |
| Final R indices [$I>2\sigma(I)$] | $R_i$ = 0.070 | 0.0302 |
| | $R_p$ = 0.140 | 0.0631 |

Table 3. Atomic coordinates and isotropic displacement parameters (Å$^2$ x 10$^3$) for Yb$_4$LiGe$_4$. The single crystal data reported are added in italics.

| Atom | Site | x | y | z | Ueq |
|------|------|------|------|------|------|
| Yb1 | 8d | 0.01487(1) | 0.09787(6) | 0.18636(1) | 8(1) |
|  |  | *0.0147(1)* | *0.0598(1)* | *0.1859(1)* | *14(1)* |
| Yb2 | 8d | 0.32746(1) | 0.12653(7) | 0.17319(1) | 9(1) |
|  |  | *0.3264(1)* | *0.1266(1)* | *0.1741(1)* | *13(1)* |
| Li | 4c | 0.0188(7) | 0.25 | 0.505(6) | 12(2) |
|  |  | *0.0158(2)* | *0.25* | *0.519(2)* | *7(3)* |
| Ge1 | 4c | 0.2821(4) | 0.25 | 0.8676(4) | 7(1) |
|  |  | *0.2808(2)* | *0.25* | *0.8690(2)* | *14(1)* |
| Ge2 | 4c | 0.0228(5) | 0.25 | 0.0931(4) | 12(1) |
|  |  | *0.0181(2)* | *0.25* | *0.0959(2)* | *14(1)* |
| Ge3 | 8d | 0.1632(4) | 0.03655(1) | 0.4664(3) | 10(1) |
|  |  | *0.1628(1)* | *0.0373(1)* | *0.4642(1)* | *15(1)* |

**Figures**

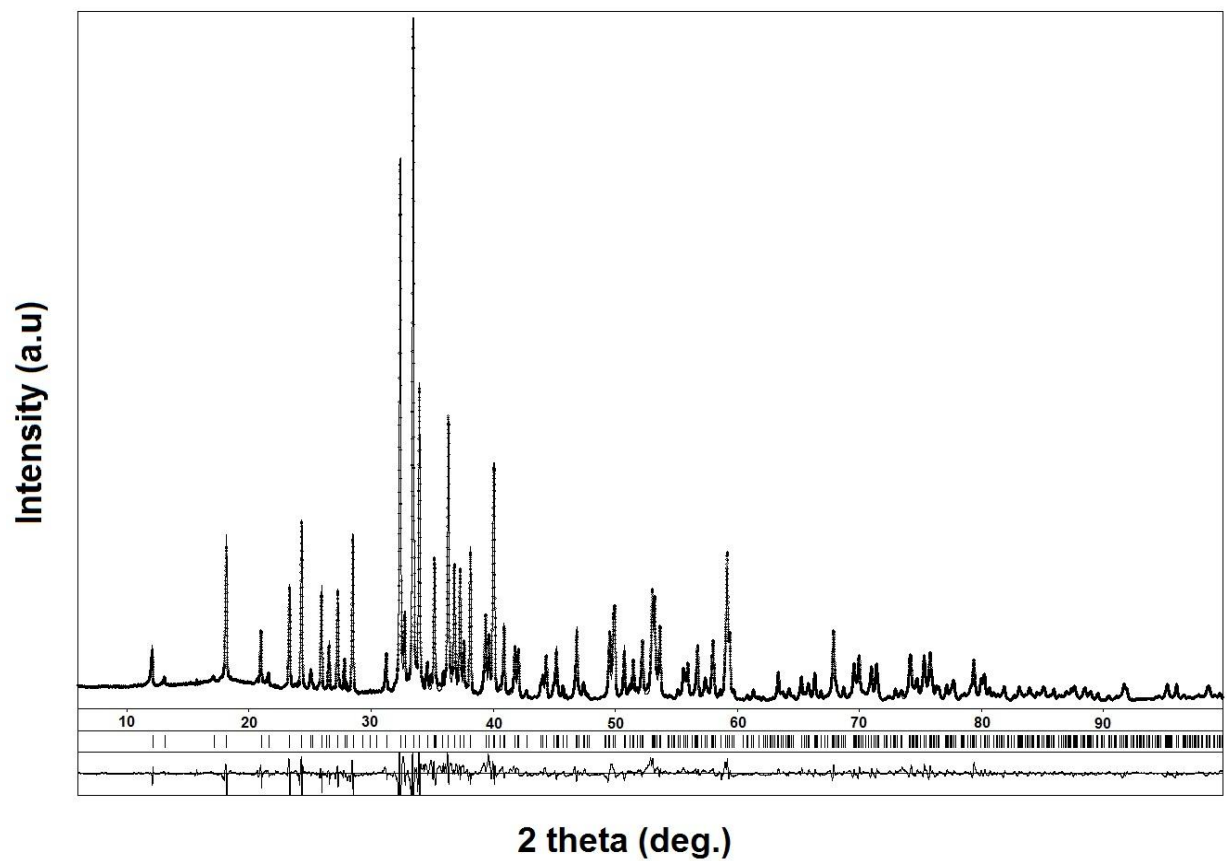

Figure 1. X-ray powder pattern measured at room temperature (+) and Reitveld fit (-) of Yb$_4$LiGe$_4$.

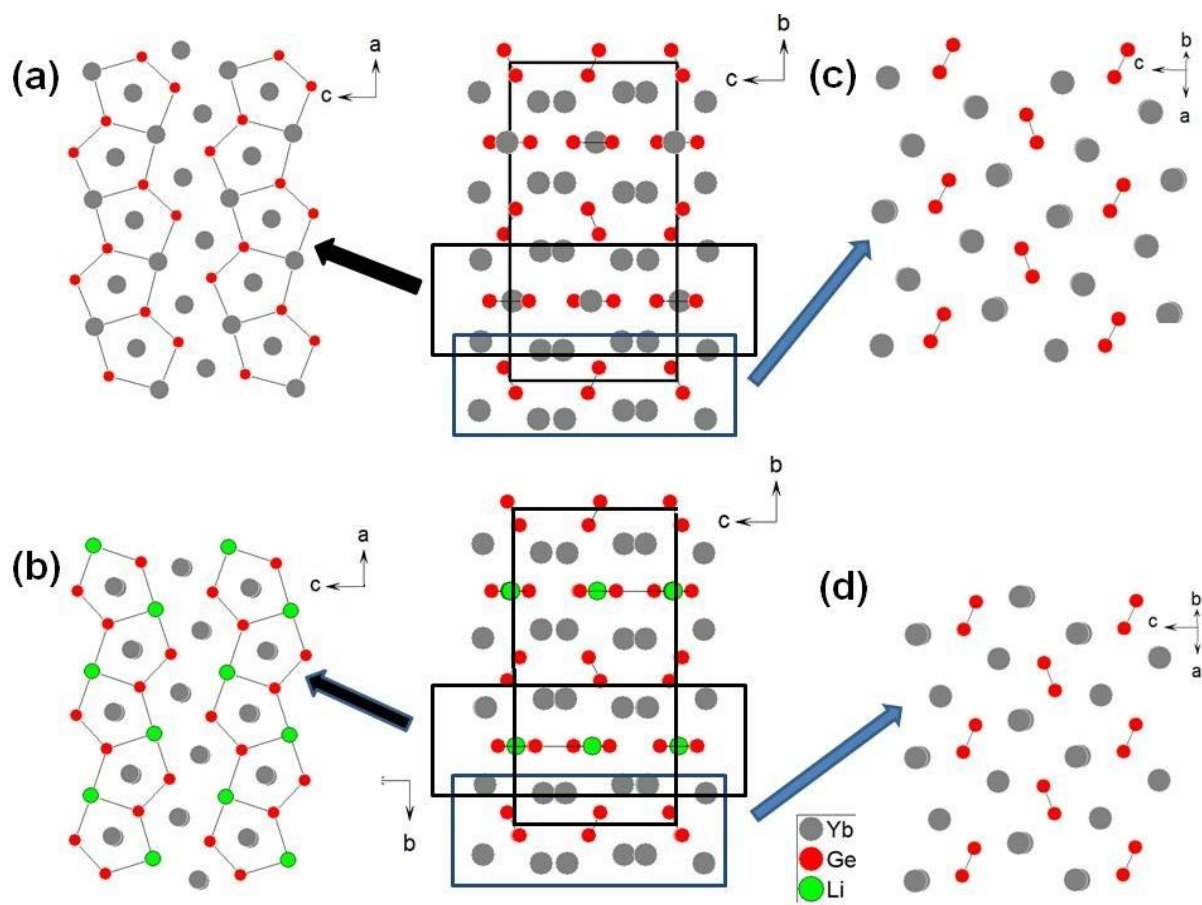

Figure 2. The two different segments of the crystal structures of $Yb_5Ge_4$ (top) and $Yb_4LiGe_4$ (bottom) within $0.05 < y < 0.45$ (a and b, segment 1) and $-0.22 < y < 0.22$ (c and d, segment 2).

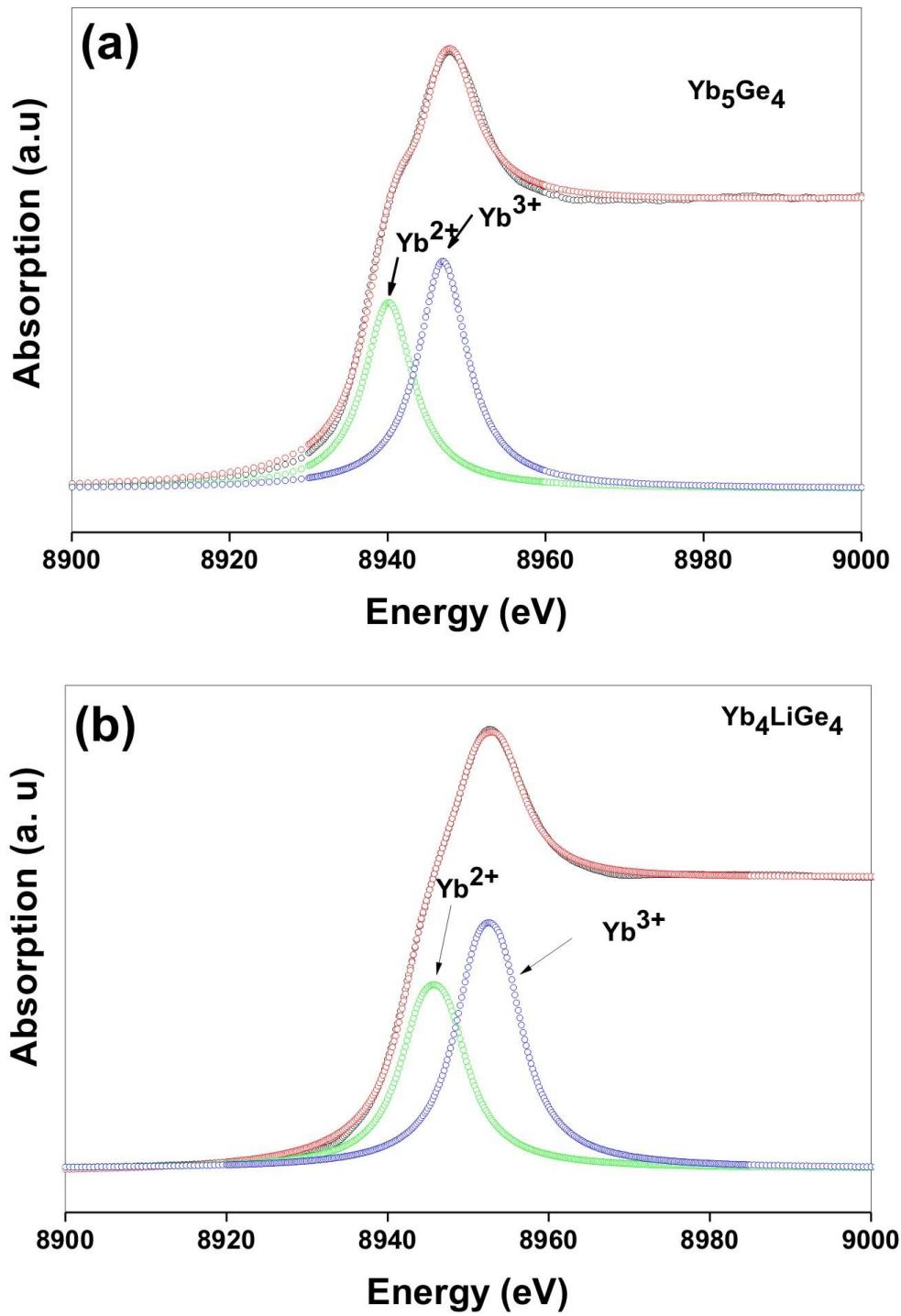

Figure 3. Yb-$L_{III}$ X-ray absorption spectra at 300 K for $Yb_5Ge_4$ (a) and $Yb_4LiGe_4$ (b) are shown in red circles. The fitted curves for $Yb^{2+}$ and $Yb^{3+}$ are represented as green and blue circles, respectively.

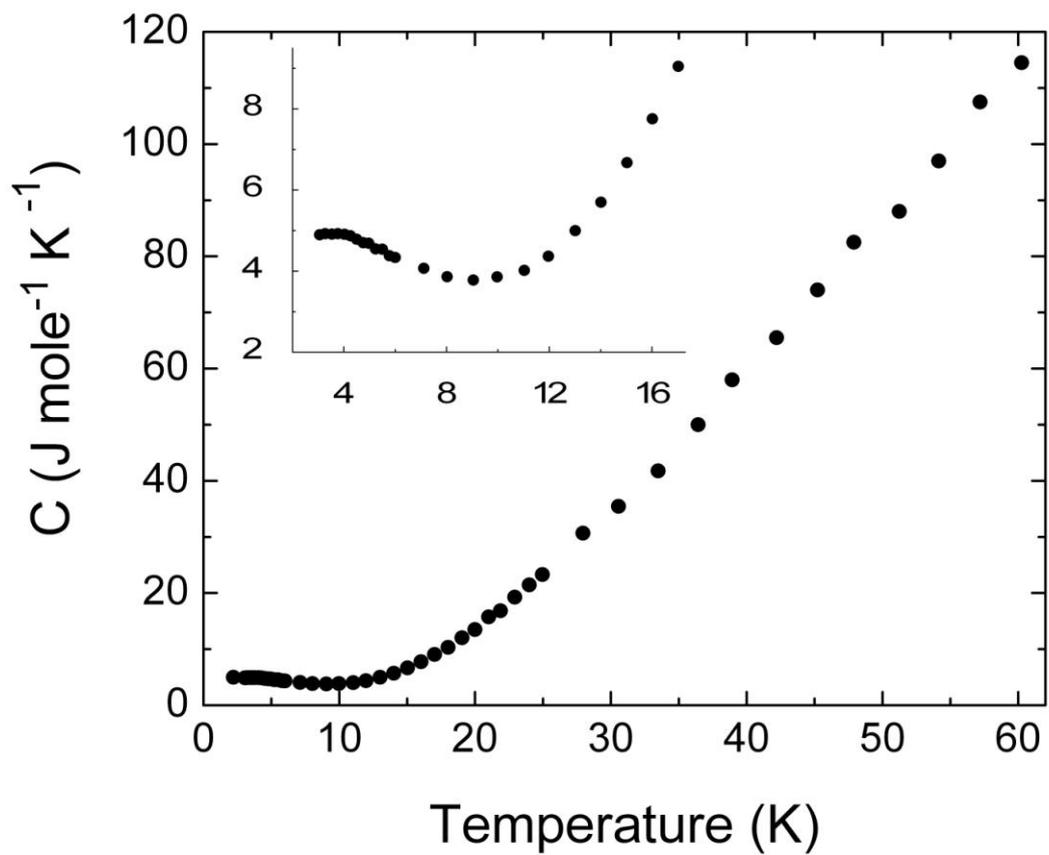

Figure 4. Specific heat of Yb$_4$LiGe$_4$ versus temperature. The low-temperature increase shown as plateau near 3 K in inset.

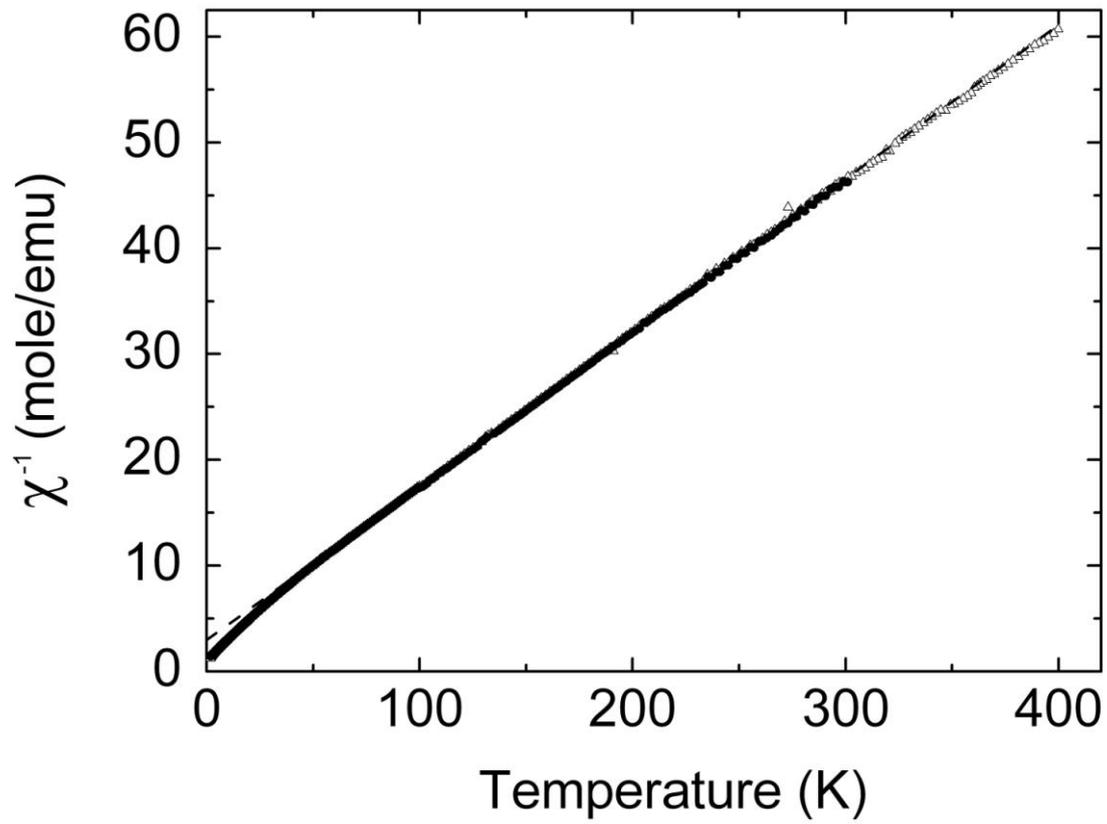

Figure 5. Inverse DC magnetic susceptibility of $Yb_4LiGe_4$ measured in fields of 1 kG and 10 kG versus temperature.

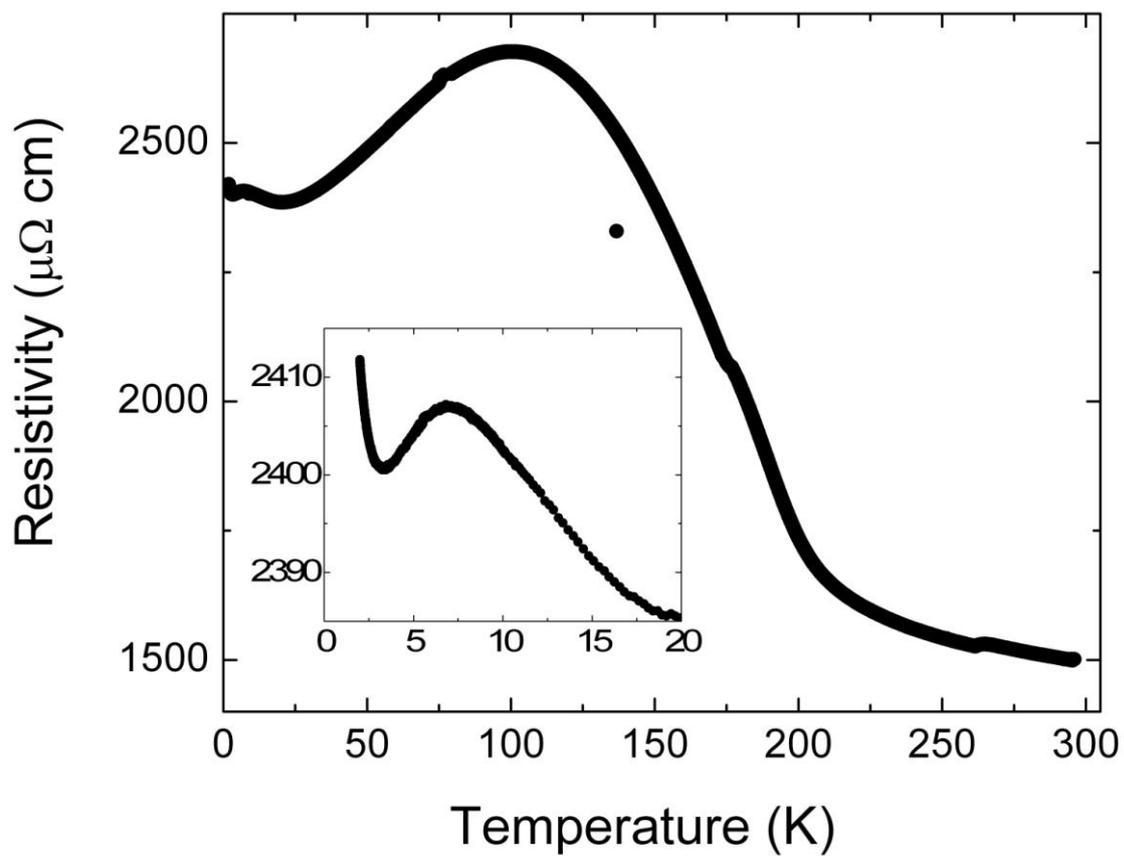

Figure 6. AC resistivity of $Yb_4LiGe_4$ versus temperature. The low temperature behavior is shown in the inset.

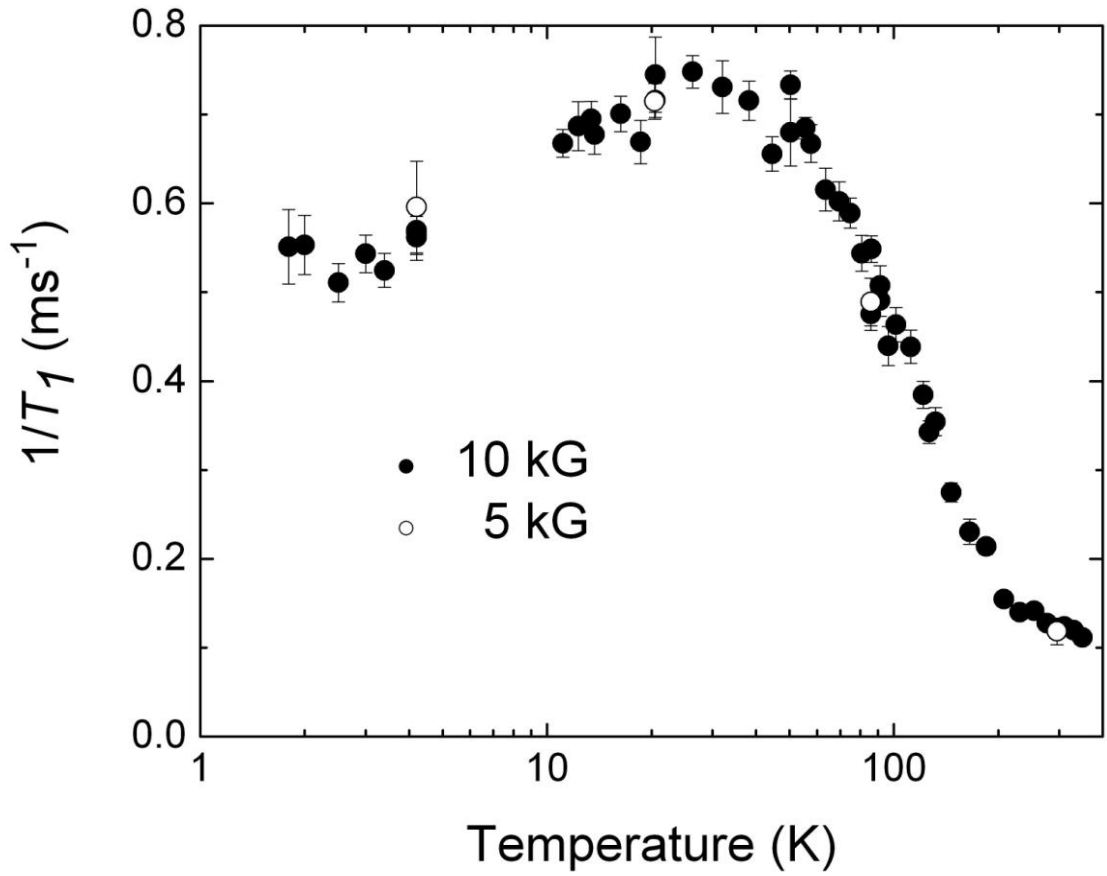

Figure 7. $^7$Li spin-lattice relaxation rate of Yb$_4$LiGe$_4$ versus temperature.

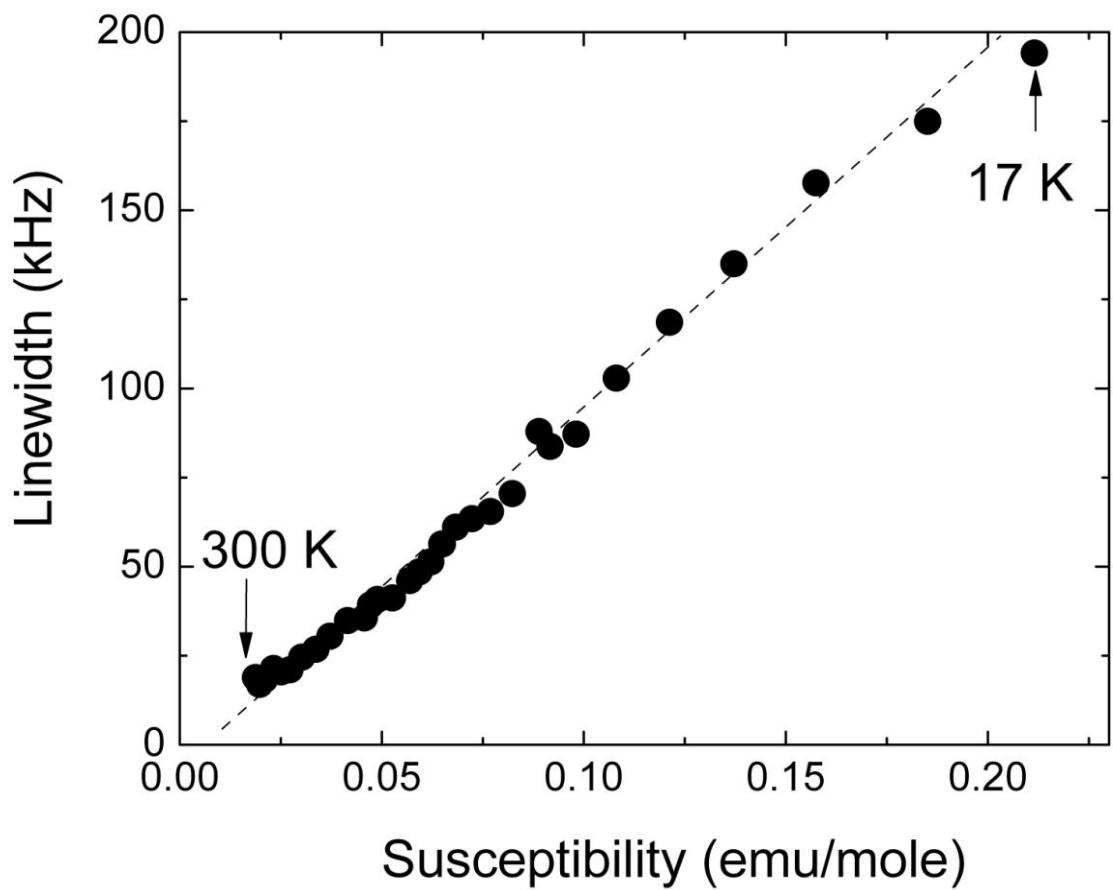

Figure 8. $^7$Li line width of $Yb_4LiGe_4$, obtained by Fourier transform of the pulse echo, plotted versus DC susceptibility in an applied magnetic field of 1 T.

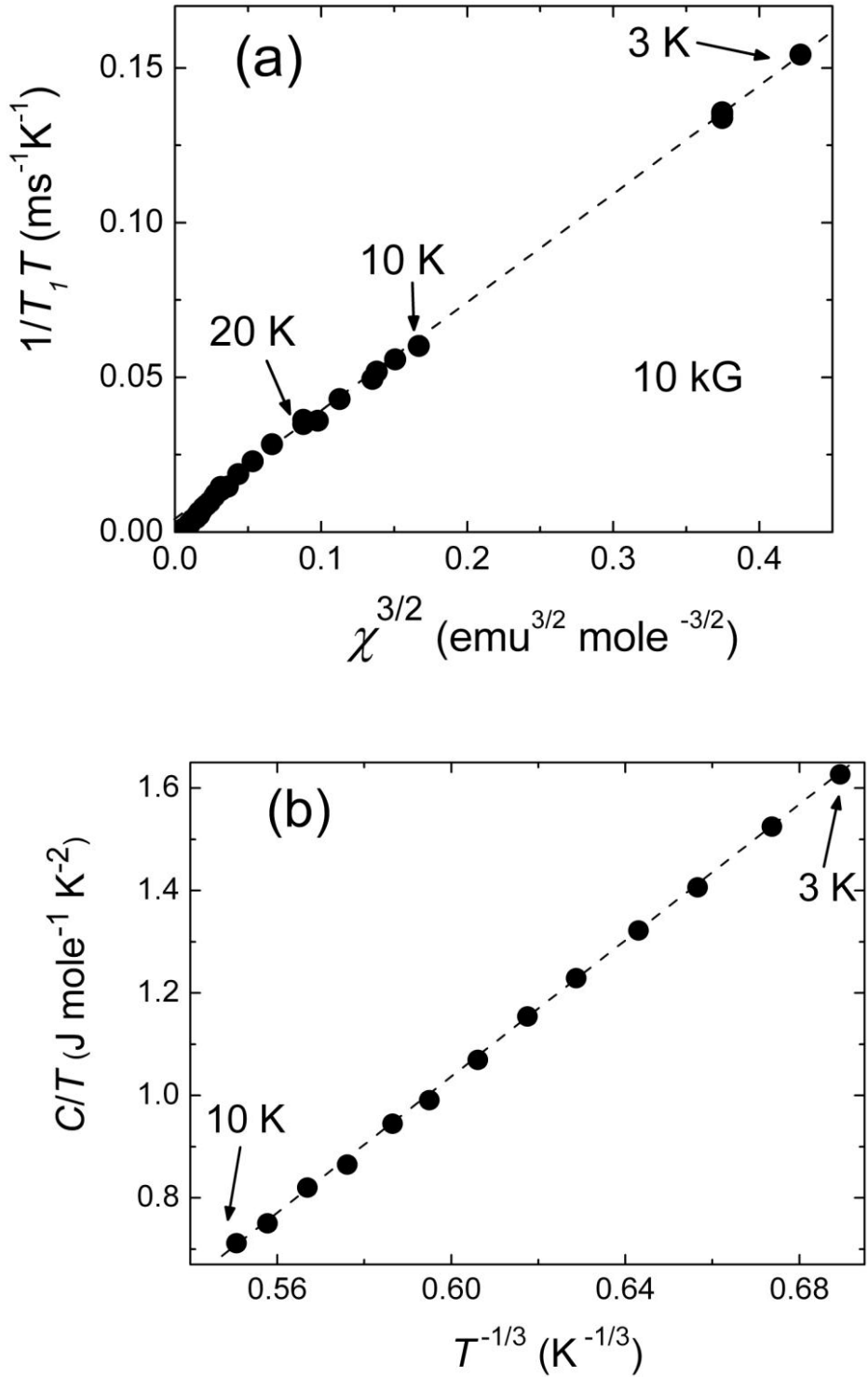

Figure 9. Analysis of the (a) $^7$Li spin-lattice relaxation and (b) specific heat of $Yb_4LiGe_4$, highlighting critical behavior which appears to follow the Ishigaki and Moriya prediction for two-dimensional system near to a low-temperature ferromagnetic phase transition described in Ref. 47.